\documentstyle[amstex,amssymb,preprint,aps]{revtex}
\tightenlines 
\begin{document}
\title{Atomic thickness hybrid F/S/F structures}
\author{A. Buzdin and M. Daumens}
\address{{\it Condensed Matter Group Theory, CPMOH,}\\
{\it \ Universit\'{e} Bordeaux I, CNRS-UMR 5798}\\
33405 TALENCE , FRANCE}
\maketitle

\begin{abstract}
We propose an exactly solvable model to describe the properties of atomic
thickness hybrid ferromagnet-superconductor-ferromagnet (F/S/F) structures.
We show that the superconducting critical temperature is always higher for
antiparallel orientation of the ferromagnetic moments. However at low
temperature the superconducting gap occurs to be larger for parallel
orientation of the ferromagnetic moments. This leads to a peculiar
temperature dependence of the proximity effect in (F/S/F) structures.
\end{abstract}

\bigskip


\section{Introduction}

In recent years the superconductor-ferromagnet (S/F) hybrid structures
attract a steadily growing interest. The actual progress in the preparation
of S/F/S junctions permits to observe the transition from normal 0-junction
to the so-called $\pi $-junction with the change of the thickness of the
ferromagnetic layer \cite{Kontos2,Blum}. Such a behavior has been predicted
in \cite{Panyukov82,Kuprianov91}\ and it is related to the oscillations of a
superconducting order parameter in the ferromagnetic layer. Another
manifestation of the spin-dependent transport in S/F hybrid structures is
the dependence of the critical temperature of metallic F/S/F sandwiches on
the mutual orientation of ferromagnetic moments of the outer layers. On the
basis of Usadel equations, applicable in the dirty limit, it has been
demonstrated in \cite{Ryzhanova99,Tagirov,Baladie1,Baladie2}, that the
antiparallel orientation of the ferromagnetic moments is more favorable for
superconductivity, i.e. it corresponds to the higher superconducting
transition temperature and to the larger amplitude of the superconducting
order parameter. Recently this effect has been observed on experiment \cite
{Gu}. Note also that for the first time the coupling between ferromagnets
through a superconducting layer has been treated theoretically in \cite
{DeGennes66} for the case of ferromagnetic insulators and observed on
experiment in \cite{Deutscher69,Hauser69}.

At the same time in \cite{Melin01,Melin02}, a microscopic multiterminal
model for S-F hybrid structure at $T=0$ has been proposed,\ where the
ferromagnet was described by different density of states for opposite
orientations of the magnetic moment. Then, it has been deduced that the
superconducting gap is less influenced by the ferromagnetism for the case of
parallel orientation of the F electrodes. Hence at $T=0,$ the parallel
orientation of the ferromagnetic moments appears more favorable for
superconductivity.

In the present work we analyze the properties of the microscopic model
system comprising three coupled atomic-scale layers: one superconducting
layer in between of two ferromagnetic layers, which can have parallel or
antiparallel orientation of the magnetic moments. This model may be relevant
for the description of layered superconductors like RuSr$_{2}$GdCu$_{8}$,
where superconducting and magnetic layers alternate \cite{McLaughlin} as
well as artificial S/F structures obtained by molecular beam epitaxy. Such a
model can be solved exactly and so we may verify that the results for the
critical temperature for a diffusive regime of electrons motion, obtained in 
\cite{Ryzhanova99,Tagirov,Baladie1,Baladie2}, are also qualitatively
applicable for a clean atomic layered system. We demonstrate that in all
cases the antiparallel alignment of ferromagnetic moments corresponds to the
higher transition temperature to the superconducting state. However at low
temperature the situation may be different and as it follows from our
analysis, the superconducting gap for parallel alignments of ferromagnetic
moments increases faster with the decrease of the temperature comparing with
the antiparallel alignment. At some temperature $T^{\ast }$ the gaps for
parallel and antiparallel alignment coincides and at $T<T^{\ast }$ the gap
is higher for parallel alignment. \ So at $T=0$ the parallel alignment is
more favorable for superconductivity while near $T_{c}$ the situation is
inverse. This result is in accordance with \cite{Melin01,Melin02}, where the
superconducting gap have been calculated at $T=0$ in the framework of quite
different model of bulk superconductor in the contact with small
ferromagnetic electrodes.

\section{Spin-orientation-dependence of superconducting transition
temperature}

Adopting the model of \cite{Andreev91}, we consider a three layers system
with one superconducting layer between two ferromagnetic layers, see Fig. 1.
It is supposed that the coupling between layers is realized via the transfer
integral $t$, which is relatively small ($t\lesssim T_{c})$, so the
superconductivity can coexist with ferromagnetism in the adjacent layers.
Introducing the notations $\varphi _{\sigma }^{+}$ and $\eta _{\sigma }^{+}$
for electrons creation operators in F layers and $\psi _{\sigma }^{+}$ in S
layer, the Hamiltonian of the system can be written as 
\[
H=H_{0}+H_{\psi }+H_{\varphi \psi }+H_{\eta \psi } 
\]
\begin{eqnarray}
H_{0} &=&\sum_{p,\sigma }\left[ \xi _{\sigma }^{\varphi }(p)\varphi _{\sigma
}^{+}(p)\varphi _{\sigma }(p)+\xi (p)\psi _{\sigma }^{+}(p)\psi _{\sigma
}(p)+\xi _{\sigma }^{\eta }(p)\eta _{\sigma }^{+}(p)\eta _{\sigma }(p)\right]
,  \label{Hamiltonian} \\
H_{\psi } &=&\sum_{p}\left[ \Delta ^{\ast }\psi _{\downarrow }(p)\psi
_{\uparrow }(-p)+\Delta \psi _{\uparrow }^{+}(p)\psi _{\downarrow }^{+}(-p)%
\right] ,  \nonumber \\
H_{\varphi \psi } &=&t\sum_{p,\sigma }\left[ \psi _{\sigma }^{+}(p)\varphi
_{\sigma }(p)+\varphi _{\sigma }^{+}(p)\psi _{\sigma }(p)\right] ,  \nonumber
\\
H_{\psi \eta } &=&t\sum_{p,\sigma }\left[ \psi _{\sigma }^{+}(p)\eta
_{\sigma }(p)+\eta _{\sigma }^{+}(p)\psi _{\sigma }(p)\right] ,  \nonumber
\end{eqnarray}
where $H_{0}$ describes the free electrons motion in F layers with the
spectra $\xi _{\sigma }^{\varphi }(p)$ and $\xi _{\sigma }^{\eta }(p)$
respectively, and with the spin-independent spectrum $\xi (p)$ in S layer.
The BCS pairing in the middle S layer is treated in $H_{\psi }$ in the mean
field approximation \cite{Abrikosov} and the coupling between neighboring
layers via the transfer integral $t$ is described by $H_{\varphi \psi }$ and 
$H_{\psi \eta }$. Note that the electrons spectra in (\ref{Hamiltonian}) are
calculated from the Fermi energy. The superconducting order parameter
satisfies the usual self-consistency equation 
\begin{equation}
\Delta ^{\ast }=\left| \lambda \right| \int \left\langle \psi _{\uparrow
}^{+}(p)\psi _{\downarrow }^{+}(-p)\right\rangle \frac{d^{2}p}{\left( 2\pi
\right) ^{2}},  \label{selcons-eq}
\end{equation}
where, $\lambda $ is the Cooper pairing constant which is assumed to be non
zero in S layers only. Introducing in the usual way \cite{Abrikosov} the\
Green functions 
\begin{eqnarray*}
G_{\alpha \beta }^{\varphi } &=&-<T_{\tau }(\varphi _{\alpha }\psi _{\beta
}^{+})>\;,\;\;F_{\alpha \beta }^{\varphi +}=<T_{\tau }(\varphi _{\alpha
}^{+}\psi _{\beta }^{+})>\;, \\
G_{\alpha \beta } &=&-<T_{\tau }(\psi _{\alpha }\psi _{\beta
}^{+})>\;,\;\;F_{\alpha \beta }^{+}=<T_{\tau }(\psi _{\alpha }^{+}\psi
_{\beta }^{+})>\;, \\
G_{\alpha \beta }^{\eta } &=&-<T_{\tau }(\eta _{\alpha }\psi _{\beta
}^{+})>\;,\;\;F_{\alpha \beta }^{\eta +}=<T_{\tau }(\eta _{\alpha }^{+}\psi
_{\beta }^{+})>\;,.
\end{eqnarray*}
and writing the corresponding equations for the Green functions, we finally
obtain the following exact expressions for the Green functions in S layer: 
\begin{eqnarray}
G_{\uparrow \uparrow } &=&\frac{b^{\ast }}{a\,b^{\ast }+\left| \Delta
\right| ^{2}}\;;\;\;G_{\downarrow \downarrow }=\frac{a^{\ast }}{a^{\ast
}\,b+\left| \Delta \right| ^{2}},  \label{sol} \\
F_{\downarrow \uparrow }^{+} &=&\frac{\Delta ^{\ast }}{a\,b^{\ast }+\left|
\Delta \right| ^{2}}\;;\;\;F_{\uparrow \downarrow }^{+}=-\frac{\Delta ^{\ast
}}{a^{\ast }b+\left| \Delta \right| ^{2}},  \nonumber
\end{eqnarray}
where 
\begin{eqnarray*}
a &=&i\omega -\xi -t^{2}\left[ \frac{1}{i\omega -\xi _{\uparrow }^{\varphi }}%
+\frac{1}{i\omega -\xi _{\uparrow }^{\eta }}\right] , \\
b &=&i\omega -\xi -t^{2}\left[ \frac{1}{i\omega -\xi _{\downarrow }^{\varphi
}}+\frac{1}{i\omega -\xi _{\downarrow }^{\eta }}\right] .
\end{eqnarray*}
The self-consistency equation for superconducting order parameter is now
written as 
\[
\Delta ^{\ast }=\left| \lambda \right| T\sum_{\omega }\int F_{\downarrow
\uparrow }^{+}\frac{d^{2}p}{\left( 2\pi \right) ^{2}}. 
\]
To calculate the critical temperature of the superconducting transition it
is sufficient to know the anomalous Green function $F_{\downarrow \uparrow
}^{+}$ in the linear approximation on $\Delta ^{\ast }.$ Then it is
convenient to write the linearized self-consistency equation for $T_{c}$ in
the following form \cite{Abrikosov} 
\begin{equation}
\ln \left( \frac{T_{c0}}{T_{c}}\right) =2T_{c}\sum_{\omega >0}\left[ \frac{%
\pi }{\omega }-\int 
%TCIMACRO{\func{Re}}%
%BeginExpansion
\mathop{\rm Re}%
%EndExpansion
(f)\,d\xi \ \right] ,  \label{eq for Tc}
\end{equation}
where we have defined a reduced function $f=1/(a^{\ast }b)$ and $T_{c0}$ is
the bare mean-field critical temperature of the central S layer in the
absence of the proximity effect (i.e. for $t=0)$.

In the case when both ferromagnetic layers are equivalent we may distinguish
two different situations

either the {\em parallel (P) orientation }of the magnetic moments, \ where $%
\xi _{\uparrow }\equiv \xi _{\uparrow }^{\varphi }=\xi _{\uparrow }^{\eta }$
and $\xi _{\downarrow }\equiv \xi _{\downarrow }^{\varphi }=\xi _{\downarrow
}^{\eta }$,

or the {\em anti-parallel (AP) orientation, }where $\xi _{\uparrow }\equiv
\xi _{\uparrow }^{\varphi }=\xi _{\downarrow }^{\eta }$ and $\xi
_{\downarrow }\equiv \xi _{\downarrow }^{\varphi }=\xi _{\uparrow }^{\eta }$.

Firstly we consider the situation when the electron dispersion spectra in
the ferromagnet\ differs only by the exchange field $h$ from the electron
spectrum in S layer i.e. $\xi _{\uparrow }=\xi -h$ and $\xi _{\downarrow
}=\xi +h$

Performing in (\ref{eq for Tc}) the integration over energy $\xi $ and
summation over Matsubara's frequencies $\omega =(2k+1)\pi T_{c}$, we finally
obtain 
\begin{eqnarray}
\ln \frac{T_{c0}}{T_{cP}} &=&\frac{1}{\pi ^{2}T_{cP}^{2}}\frac{2t^{2}}{%
8t^{2}+h^{2}}\left\{ h^{2}\Phi _{1}\left( \frac{h}{2\pi T_{cP}}\right)
\right. +  \label{eq for Ta and Tp} \\
&&\left. 4t^{2}\left[ \Phi _{1}\left( \frac{\sqrt{8t^{2}+h^{2}}+h}{2\pi
T_{cP}}\right) +\Phi _{1}\left( \frac{\sqrt{8t^{2}+h^{2}}-h}{2\pi T_{cP}}%
\right) \right] \right\} ,  \nonumber
\end{eqnarray}
\[
\ln \frac{T_{c0}}{T_{cAP}}=\frac{1}{\pi ^{2}T_{cAP}^{2}}\frac{2t^{2}}{%
2t^{2}+h^{2}}\left[ h^{2}\Phi _{1}\left( \frac{\sqrt{2t^{2}+h^{2}}}{2\pi
T_{cAP}}\right) +2t^{2}\Phi _{1}\left( \frac{\sqrt{2t^{2}+h^{2}}}{\pi T_{cAP}%
}\right) \right] , 
\]
where $T_{cP}$ ($T_{cAP}$) is the superconducting transition temperature for
parallel (antiparallel) orientation of the F-layers magnetic moments and we
define the function $\Phi _{1}$ through the Digamma function $\Psi $ as ($%
\gamma $ is Euler constant) 
\[
\Phi _{1}(x)=\frac{1}{2x^{2}}\left\{ \gamma +2\ln (2)+\frac{1}{2}\left[ \Psi
\left( \frac{1+ix}{2}\right) +\Psi \left( \frac{1-ix}{2}\right) \right]
\right\} . 
\]
Using (\ref{eq for Ta and Tp}) we can determine the relative variation of
critical temperature $\delta T=(T_{c}-T_{c0})/T_{c0}$ in the two following
limits :

{\bf i) }in the limit{\bf \ }$\frac{\sqrt{t^{2}+h^{2}}}{T_{c}}\rightarrow 0$
up to the fourth order over $t$ we have: 
\begin{eqnarray*}
\delta T_{P} &\simeq &-\frac{7\zeta (3)}{\left( 2\pi T_{c0}\right) ^{2}}%
t^{2}+\frac{1}{\left( 2\pi T_{c0}\right) ^{4}}\left[ \left( 62\zeta (5)-%
\frac{147}{2}\zeta (3)^{2}\right) t^{4}+\frac{31}{4}\zeta (5)t^{2}h^{2}%
\right] + \\
&&+\frac{1}{\left( 2\pi T_{c0}\right) ^{6}}\left[ \left( \frac{1085}{16}%
\zeta (3)\zeta (5)-\frac{889}{8}\zeta (7)\right) t^{4}h^{2}-\frac{127}{16}%
\zeta (7)t^{2}h^{4}\right] ,
\end{eqnarray*}
\begin{equation}
\delta T_{P}-\delta T_{AP}\simeq -\frac{1397\zeta (7)}{512\pi ^{6}}\frac{%
t^{4}h^{2}}{T_{c0}^{6}}\simeq -0.0029\frac{t^{4}h^{2}}{T_{c0}^{2}}.
\label{diff Tc for small h}
\end{equation}
The last result, the difference $\delta T_{P}-\delta T_{AP},$ was already
found for S/F multilayer in \cite{Baladie1} (after correcting a sign
mistake).

{\bf ii) }in the limit{\bf \ }$\frac{\sqrt{t^{2}+h^{2}}}{T_{c}}\rightarrow
\infty $ and for $t<<h$ we obtain : 
\[
\delta T_{P}\simeq -4\frac{t^{2}}{h^{2}}\left[ \gamma +\ln \left( \frac{h}{%
\pi T_{c0}}\right) +\frac{7\zeta (3)}{4\pi ^{2}}\frac{t^{2}}{T_{c0}^{2}}%
\right] , 
\]
\[
\delta T_{A}\simeq -4\frac{t^{2}}{h^{2}}\left[ \gamma +\ln \left( \frac{h}{%
\pi T_{c0}}\right) \right] , 
\]
\begin{equation}
\delta T_{P}-\delta T_{AP0}\simeq -\frac{7\zeta (3)}{4\pi ^{2}}\frac{t^{4}}{%
h^{2}T_{c0}^{2}}\simeq -0.21\frac{t^{4}}{h^{2}T_{c0}^{2}}.
\label{diff Tc for large h}
\end{equation}
Also the difference of the critical temperatures $\delta T_{P}-\delta T_{AP}$
coincides with that of the S/F multilayer \cite{Baladie1}.

We see that in all cases the difference between critical temperatures for
parallel and antiparallel alignment is proportional to $t^{4}$, while the
decrease of the critical temperature itself is proportional to $t^{2}.$ This
fact demonstrates that the spin-orientation dependence of $T_{c}$ is related
with \ a rather subtle interference effect between electrons coming from
ferromagnetic layers.

Now we consider the case of a ferromagnetic half-metal, which we may model
in (\ref{sol}) by $\xi _{\downarrow }=\xi +h$ and $\xi _{\uparrow
}\rightarrow \infty ,\ $that corresponds to a zero density of sates for the
electrons with spin orientation along magnetic moment in ferromagnetic
layers. Note that the limit $\xi _{\uparrow }\rightarrow \infty \ $is
equivalent to take the transfer integral for electrons with spin up
orientation equal to zero. This may be simply demonstrated\ from the initial
equations for the Green functions$.$ Performing integration over $\xi $ and
summation over Matsubara frequencies in the self-consistency equation, we
can deduce the relative temperature variation $\delta T=(T_{c}-T_{c0})/T_{c0}
$ at the order up to $t^{4}$, in terms of the dimensionless field $\widehat{h%
}=h/(2\pi T_{c0})$%
\begin{eqnarray*}
\delta T_{P} &=&-\Phi _{1}\left( \widehat{h}\right) \frac{t^{2}}{\pi
^{2}T_{c0}^{2}} \\
&&+\left\{ \frac{1}{2}\Phi _{1}\left( \widehat{h}\right) ^{2}-2\Phi
_{1}\left( \widehat{h}\right) \Phi _{2}\left( \widehat{h}\right) -\frac{1}{%
\widehat{h}^{2}}\left[ \frac{7}{16}\zeta (3)-\frac{3}{2}\Phi _{1}\left( 
\widehat{h}\right) +\Phi _{2}\left( \widehat{h}\right) \right] \right\} 
\frac{t^{4}}{\pi ^{4}T_{c0}^{4}},
\end{eqnarray*}
and 
\begin{equation}
\delta T_{P}-\delta T_{AP}=-\frac{1}{2\widehat{h}^{2}}\left[ \frac{7}{8}%
\zeta (3)-\Phi _{1}\left( \widehat{h}\right) \right] \frac{t^{4}}{\pi
^{4}T_{c0}^{4}},  \label{Tc semimetal}
\end{equation}
where the function $\Phi _{1}$ was defined before and the function $\Phi _{2}
$ is
\[
\Phi _{2}(x)=\frac{1}{16i}\left[ \Psi ^{\prime }\left( \frac{1-ix}{2}\right)
-\Psi ^{\prime }\left( \frac{1+ix}{2}\right) \right] .
\]
In the limiting case $h\rightarrow 0,$which corresponds to the situation $%
h\ll T_{c0},$or simply $\xi _{\uparrow }=\xi ,$ we have
\[
\delta T_{P}=-\left[ \frac{7}{8}\zeta (3)\right] \frac{t^{2}}{\pi
^{2}T_{c0}^{2}}+\left\{ \frac{31}{64}\zeta (5)-\frac{3}{2}\left( \frac{7}{8}%
\zeta (3)\right) ^{2}\right\} \frac{t^{4}}{\pi ^{4}T_{c0}^{4}},
\]
and the critical temperature difference is 
\begin{equation}
\delta T_{P}-\delta T_{AP}=-\frac{31}{64}\zeta (5)\frac{t^{4}}{\pi
^{4}T_{c0}^{4}}\simeq -0.50\frac{t^{4}}{\pi ^{4}T_{c0}^{4}}.
\end{equation}
In the opposite limit $h\gg T_{c0},$ we find 
\[
\delta T_{P}=-\frac{1}{2\widehat{h}^{2}}\ln \left( \widehat{h}\right) \frac{%
t^{2}}{\pi ^{2}T_{c0}^{2}}-\frac{7}{16}\frac{\zeta (3)}{\widehat{h}^{2}}%
\frac{t^{4}}{\pi ^{4}T_{c0}^{4}},
\]
and 
\begin{equation}
\delta T_{P}-\delta T_{AP}=-\frac{4t^{4}}{h^{4}}\ln \left( \frac{h}{2\pi
T_{c0}}\right) .
\end{equation}
Then we may conclude that in all cases the critical temperature is higher
for the antiparallel alignment of the ferromagnetic moments.

\section{\protect\bigskip Superconducting gap at low temperatures}

In this section we demonstrate that the proximity effect at low temperature
is very special and that the superconducting gap at $T=0,$ for half-metal
and for usual ferromagnet with small exchange field, is higher for parallel
orientation in accordance with \cite{Melin01,Melin02}.

Firstly we consider the model of a half-metal with $\xi _{\uparrow
}\rightarrow \infty $ and $\xi _{\downarrow }=\xi .$ So the self consistency
equations for parallel and antiparallel orientations may be written as 
\begin{eqnarray}
\frac{1}{\left| \lambda \right| N(0)} &=&T\sum_{\omega }\int \frac{d\xi }{%
\omega ^{2}+\xi ^{2}+2t^{2}%
%TCIMACRO{\dfrac{i\omega -\xi }{i\omega +\xi }}%
%BeginExpansion
{\displaystyle{i\omega -\xi  \over i\omega +\xi }}%
%EndExpansion
+\Delta _{P}^{2}},  \nonumber \\
&=&T\sum_{\omega }\int \frac{d\xi }{\omega ^{2}+\xi ^{2}+2t^{2}%
%TCIMACRO{\dfrac{\omega ^{2}-\xi ^{2}}{\omega ^{2}+\xi ^{2}}}%
%BeginExpansion
{\displaystyle{\omega ^{2}-\xi ^{2} \over \omega ^{2}+\xi ^{2}}}%
%EndExpansion
+%
%TCIMACRO{\dfrac{t^{4}}{\omega ^{2}+\xi ^{2}}}%
%BeginExpansion
{\displaystyle{t^{4} \over \omega ^{2}+\xi ^{2}}}%
%EndExpansion
+\Delta _{AP}^{2}}
\end{eqnarray}
where $N(0)$ is the electron density of state.

In the limit of weak interlayer coupling $t<<T_{c0}$ and at $T=0,$ we obtain
the following expression for the superconducting gap for parallel
orientation $\Delta _{P}$%
\begin{equation}
\ln \left( \frac{\Delta _{0}}{\Delta _{P}}\right) =\frac{\pi }{2}%
\int_{0}^{2\pi }\ln \left[ 1+2\frac{t^{2}}{\Delta _{P}^{2}}\exp (2i\theta )%
\right] d\theta ,
\end{equation}
where $\Delta _{0}$ is the gap of an isolated S layer. Performing expansion
over $t,$ it may be demonstrated that at all order over $t^{2\text{ }}$the
corrections to $\Delta _{P}$ disappear, and $\Delta _{P}=\Delta _{0}$ .\ So
the proximity effect for the case of half-metal vanishes. It may be
understood as an impossibility of Cooper pair destruction. Indeed at $T=0$,
the disappearance of the Cooper pair in S layer means that two electrons
with opposite spins must leave it. Due to the insulating character of
neighboring F layer for some one spin orientation it becomes impossible and
so, the Cooper pair is not destroyed at all.

On the other hand, for the antiparallel alignment this argument does not
works and\ the proximity effect leads to a decrease of the gap $\Delta _{AP}$%
\begin{equation}
\ln \left( \frac{\Delta _{0}}{\Delta _{AP}}\right) =\frac{t^{4}}{\Delta
_{AP}^{4}}\left( 1+2\ln \frac{\Delta _{AP}}{t}\right) .
\end{equation}
Therefore the superconducting gap variation is $\left( \Delta _{0}-\Delta
_{AP}\right) /\Delta _{0}\thickapprox 2\left( t/\Delta _{0}\right)
^{4}ln\left( \Delta _{0}/t\right) .$ As the superconducting transition
temperature is higher for the AP case, it means that above some temperature $%
T^{\ast \text{ }}$the gap for parallel orientation becomes smaller than that
for antiparallel orientation.

To find this temperature $T^{\ast \text{ }}$in the limit $t<<T_{c0}$ we
perform an expansion over $t$ in the self-consistency equations for $\Delta
_{AP}=\Delta _{P}=\Delta ^{\ast }.$ After integration over energy $\xi $ we
have the following equation for the ratio $X=\Delta ^{\ast }/(\pi T^{\ast })$%
\begin{equation}
\sum_{k}\frac{%
2(2k+1)^{3}+5(2k+1)X^{2}-2[5(2k+1)^{2}-X^{2}][(2k+1)^{2}+X^{2}]^{1/2}}{%
(2k+1)[(2k+1)^{2}+X^{2}]^{3/2}[(2k+1)+\sqrt{(2k+1)^{2}+X^{2}}]^{4}}=0,
\end{equation}
from what we find $\Delta ^{\ast }/T^{\ast \text{ }}=$ $4.22$. Taking into
account that in the limit $t<<T_{c0}$ the temperature dependence of the
superconducting gap in the first approximation is the same as that for an
isolated S layer $\Delta _{0}(T)$, we directly find the temperature of gap
inversion $T^{\ast \text{ }}=0.41T_{c0}.$

Similar calculations can be performed for the standard ferromagnetic case
with $\xi _{\uparrow }=\xi -h$ and $\xi _{\downarrow }=\xi +h$ in the limit $%
t,h\ll T_{c0}.$ In the result we obtain the following equation for the ratio 
$X=\Delta ^{\ast }/(\pi T^{\ast })$%
\begin{gather}
\sum_{k}\left\{ \frac{-2(2k+1)^{4}-16(2k+1)^{2}X^{2}-6X^{4}}{%
(2k+1)^{2}[(2k+1)^{2}+X^{2}]^{3/2}[(2k+1)+\sqrt{(2k+1)^{2}+X^{2}}]^{6}}%
+\right.   \nonumber \\
\left. \frac{+38(2k+1)^{4}-11(2k+1)^{2}X^{2}-X^{4}}{%
(2k+1)^{3}[(2k+1)^{2}+X^{2}][(2k+1)+\sqrt{(2k+1)^{2}+X^{2}}]^{6}}\right\} =0,
\end{gather}
which gives $\Delta ^{\ast }/T^{\ast \text{ }}=3.61$, and the temperature $%
T^{\ast \text{ }}$ of the gap inversion is $T^{\ast \text{ }}=0.47T_{c0}.$

\section{Conclusion}

In the present work we have demonstrated that, in the framework of the
considered microscopic model, in all cases the superconducting transition
temperature is higher for antiparallel orientation of the ferromagnetic
moments. This is consistent with the predictions made in \cite
{Ryzhanova99,Tagirov,Baladie1,Baladie2}, on the basis of the diffusive dirty
limit model and recent experimental results \cite{Gu}. We may consider this
result as a quite general one and model independent. On the other hand the
superconducting gap at low temperature for the case of half-metal and for
usual ferromagnet with small exchange field occurs to be larger for parallel
orientation. This is in opposite with the diffusive model prediction \cite
{Baladie2}, but in accordance with the $T=0$ result\ \cite{Melin01,Melin02}
obtained in the framework of the multiterminal model for S-F hybrid
structures. It may be interesting to calculate the superconducting critical
temperature in the model \cite{Melin01,Melin02} to compare with our
predictions.

Finally we conclude that the spin-orientation dependence of the
superconductivity in F/S/F atomic layers structure may be quite different at
high and low temperature regimes. Note also that the condition of small
interlayer coupling $t<<T_{c0}$, may be crucial for the conclusion that
there is a gap inversion for parallel and antiparallel orientation with
temperature decrease. Indeed, it has been demonstrated in \cite{Houzet},
that the strength of interlayer coupling may qualitatively change the
critical temperature dependence via the exchange field.

Note also, that strictly speaking, the superconductivity is impossible in
atomic 2D system. However, our results must be qualitatively applicable for
systems consisting of several consecutive S and F layers, as well as for S/F
multilayered systems, where the fluctuations are strongly suppressed.

\bigskip

{\large ACKNOWLEDGMENTS}

We are grateful to D. Feinberg and R. Melin for useful discussions and
correspondence.

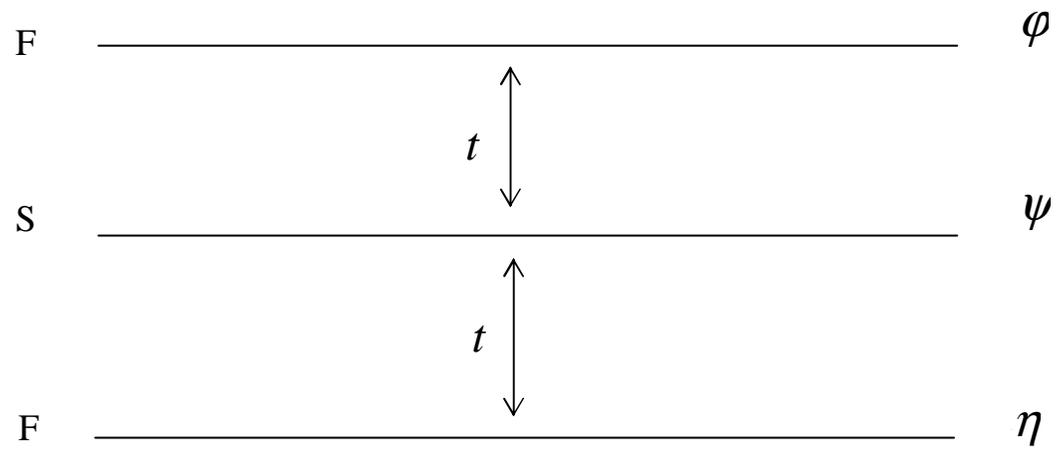
\end{document}